\documentclass[conference]{IEEEtran}
\IEEEoverridecommandlockouts
\usepackage{cite}
\usepackage{amsmath,amssymb,amsfonts}
\usepackage{algorithmic}
\usepackage{graphicx}
\usepackage{textcomp}
\usepackage{xcolor}
\usepackage{graphicx}
\usepackage{subfigure}

\def\BibTeX{{\rm B\kern-.05em{\sc i\kern-.025em b}\kern-.08em
    T\kern-.1667em\lower.7ex\hbox{E}\kern-.125emX}}
\begin{document}

\title{Reconfigurable Surface Wave Platform Using\\Fluidic Conductive Structures
}

\author{\IEEEauthorblockN{Zhiyuan Chu, Kai-Kit Wong, Kin-Fai Tong}
\IEEEauthorblockA{\textit{Department of Electronic and Electrical Engineering} \\
\textit{University College London}\\
London, United Kingdom  \\
zhiyuan.chu.18@ucl.ac.uk, kai-kit.wong@ucl.ac.uk, k.tong@ucl.ac.uk}

}
\maketitle

\begin{abstract}
Surface wave inherently has less propagation loss as it adheres to the surface and minimizes unwanted dissipation in space. Recently, they find applications in network-on-chip (NoC) communications and intelligent surface aided mobile networked communications. This paper puts forward a reconfigurable surface wave platform (RSWP) that utilizes liquid metal to produce highly energy-efficient and adaptive pathways for surface wave transmission. Our simulation results illustrate that the proposed RSWP using Galinstan can obtain a $25{\rm dB}$ gain in the electric field for a propagation distance of $35\lambda$ at $30{\rm GHz}$ where $\lambda$ denotes the wavelength. Moreover, less than $1{\rm dB}$ loss is observed even at a distance of $50\lambda$, and a pathway with right-angled turns can also be created with only a $3.5{\rm dB}$ loss at the turn.
\end{abstract}


\section{Introduction}
The fact that surface wave spreads over a 2-dimensional	(2d) space, means that it boasts superior propagation characteristics compared to space waves [1]. Surface wave has already been proposed for network-on-chip (NoC) communications [2]. The recent rise of intelligent reflecting surface (IRS) for 6G mobile communications further sees the trend of using reconfigurable surfaces to create smart radio environments [3]. The adoption of surface waves in 6G has also been advocated in [4].

Nevertheless, little is known for developing reconfigurable surface waves as envisaged in [4]. The most relevant results appeared in [5] where it was hypothesized that mechanically controlled metal bars were made to appear/disappear to create changeable surface wave pathways. Although the concept is appealing, the practicality of the approach in [5] is questionable. In this paper, we propose a reconfigurable surface wave platform (RSWP) which utilizes liquid metal bars that can be easily manipulated to form specific pathways.

\begin{figure}[]
\centering
\includegraphics[width=9cm]{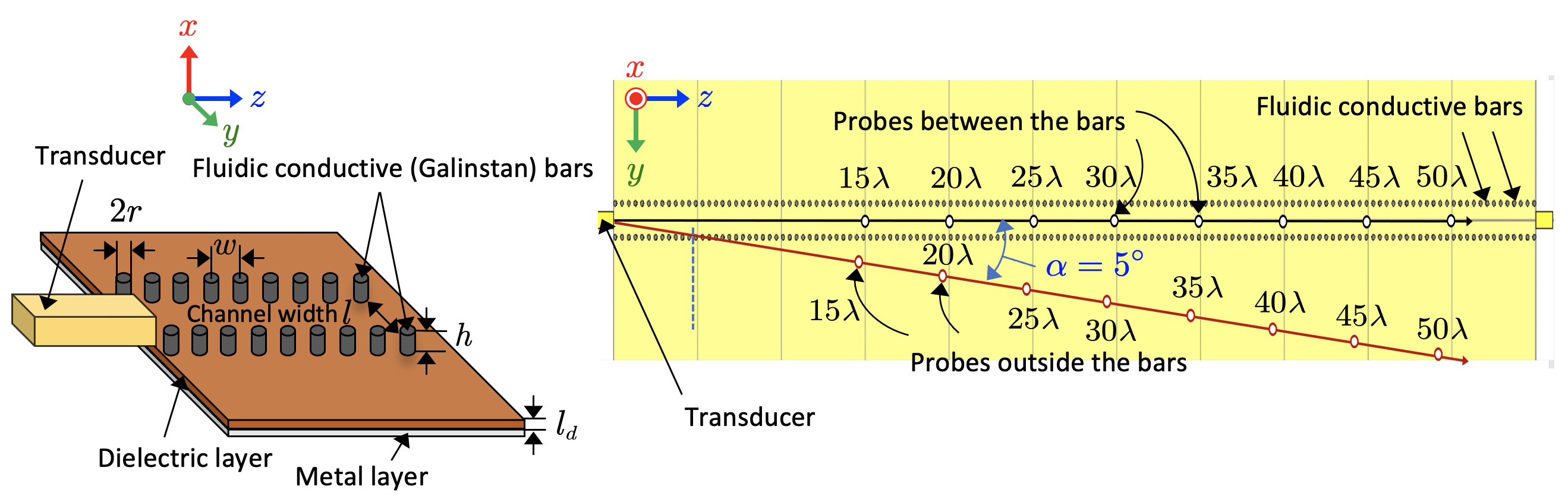}
\caption{The RSWP with two rows of bars filled with liquid metal (left) and an illustration of the probe locations for measuring the electric field (right).}\label{Models}
\end{figure}

\section{The RSWP Model}
We consider an RSWP with hollow bars evenly distributed over the entire dielectric layer and each of the bars may be filled with liquid metal on-demand. In Fig.~\ref{Models}, the RSWP model with two rows of bars filled with liquid metal is illustrated to create a straight pathway. Note that only bars filled with conductive fluid are shown although periodic cavity bars are installed all over the entire surface. The RSWP is expected to form any arbitrary 2d pathway (with sharp turns if needed) by pumping conductive fluid to fill specific cavity bars.

\section{Simulation Results}
Full-wave 3-dimensional electromagnetic simulations using CST Studio Suite 2020 were conducted. In the simulations, Galinstan was chosen as the liquid metal alloy to fill the bars when needed and Rogers RT5880 (lossy) was employed as the dielectric layer. The operating frequency was set to be $30{\rm GHz}$ and the transducer had an aperture with height of $2.84{\rm mm}$ and width of $5.89{\rm mm}$. Electric field sampling probes were added at the locations specified in Fig.~\ref{Models} where the electric field in the $x$-direction, i.e., $E_x$, was measured. The parameters selected in the simulations are summarized in TABLE \ref{tab:setup}.

\begin{table}[]
\begin{center}
\begin{tabular}{r||c}
{\bf Parameter} & {\bf Value}\\
\hline
\mbox{height of the bar}, $h$ & $5{\rm mm}$\\
\hline
\mbox{radius of the bar}, $r$ & $1{\rm mm}$\\
\hline
\mbox{center-to-center separation between bars}, $w$ & $4{\rm mm}$\\
\hline
\mbox{channel width}, $l$ & $6{\rm mm}$\\
\hline
\mbox{relative permittivity of the dielectric layer}, $\varepsilon_r$ & $2.2$\\
\hline
\mbox{thickness of the dielectric layer}, $l_d$ & $1{\rm mm}$\\
\hline
\mbox{inductive surface impedance} & $j130\Omega$\\
\hline
\mbox{conductivity for Galinstan}, $\sigma_{\rm g}$ & $3.46\times 10^6~{\rm Sm}^{-1}$\\
\hline
\mbox{conductivity for copper}, $\sigma_{\rm c}$ & $59.6\times 10^6~{\rm Sm}^{-1}$
\end{tabular}
\end{center}
\caption{The parameters used in the simulations.}\label{tab:setup}
\vspace{-5mm}
\end{table}


\begin{figure}[t]
\centering
\subfigure[]{
\includegraphics[height=5cm]{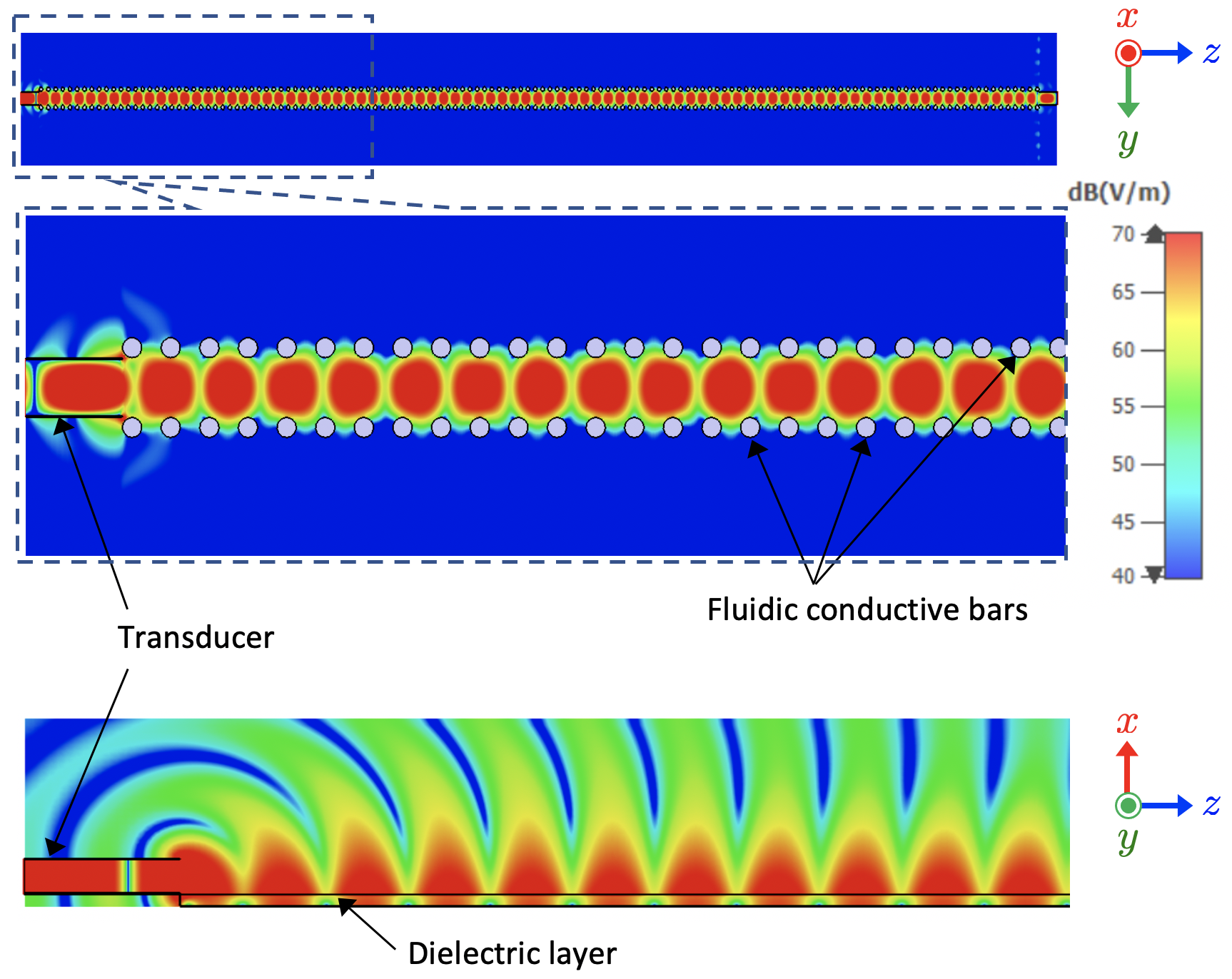}
}
\subfigure[]{
\includegraphics[height=5cm]{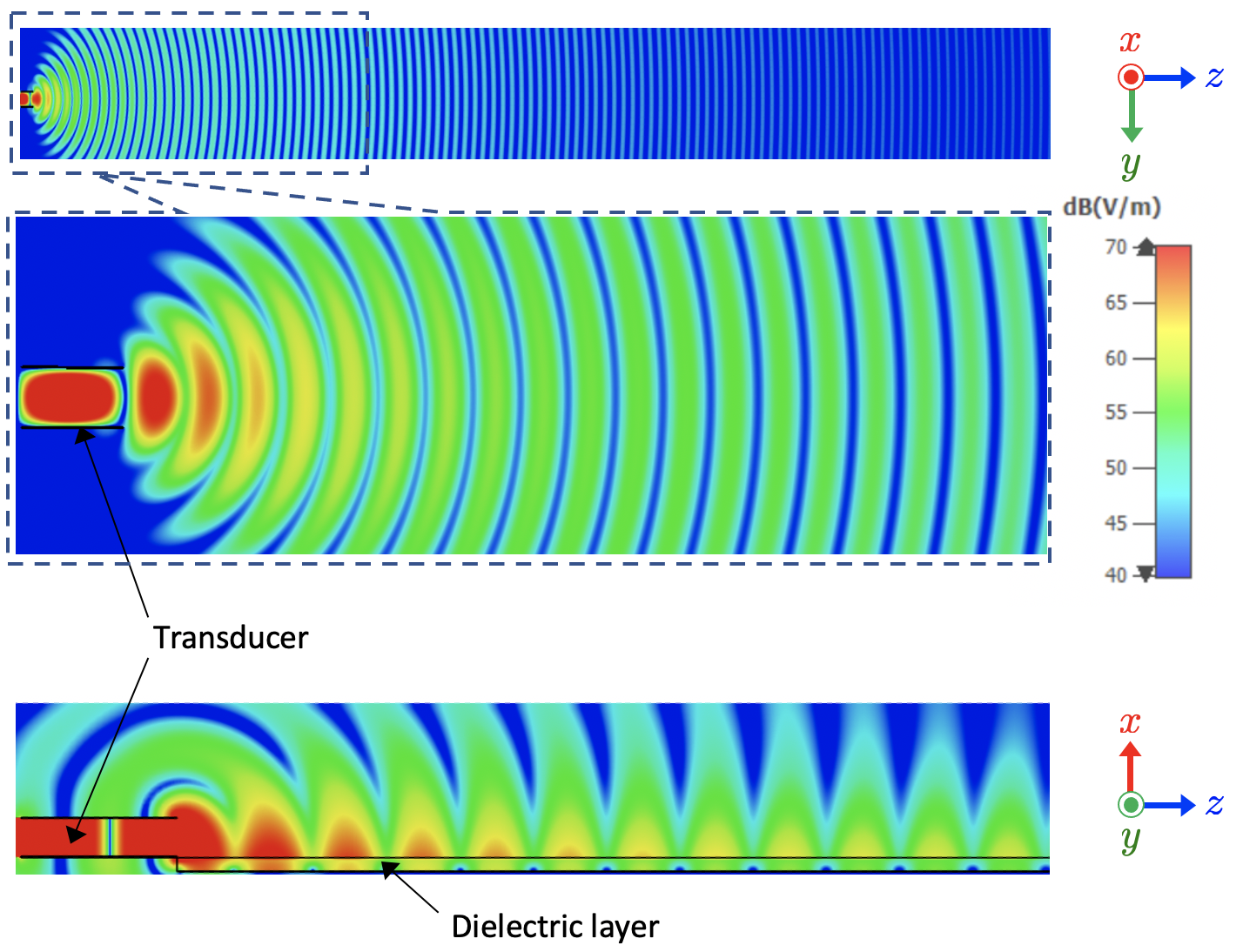}
}
\subfigure[]{
\includegraphics[width=8.8cm]{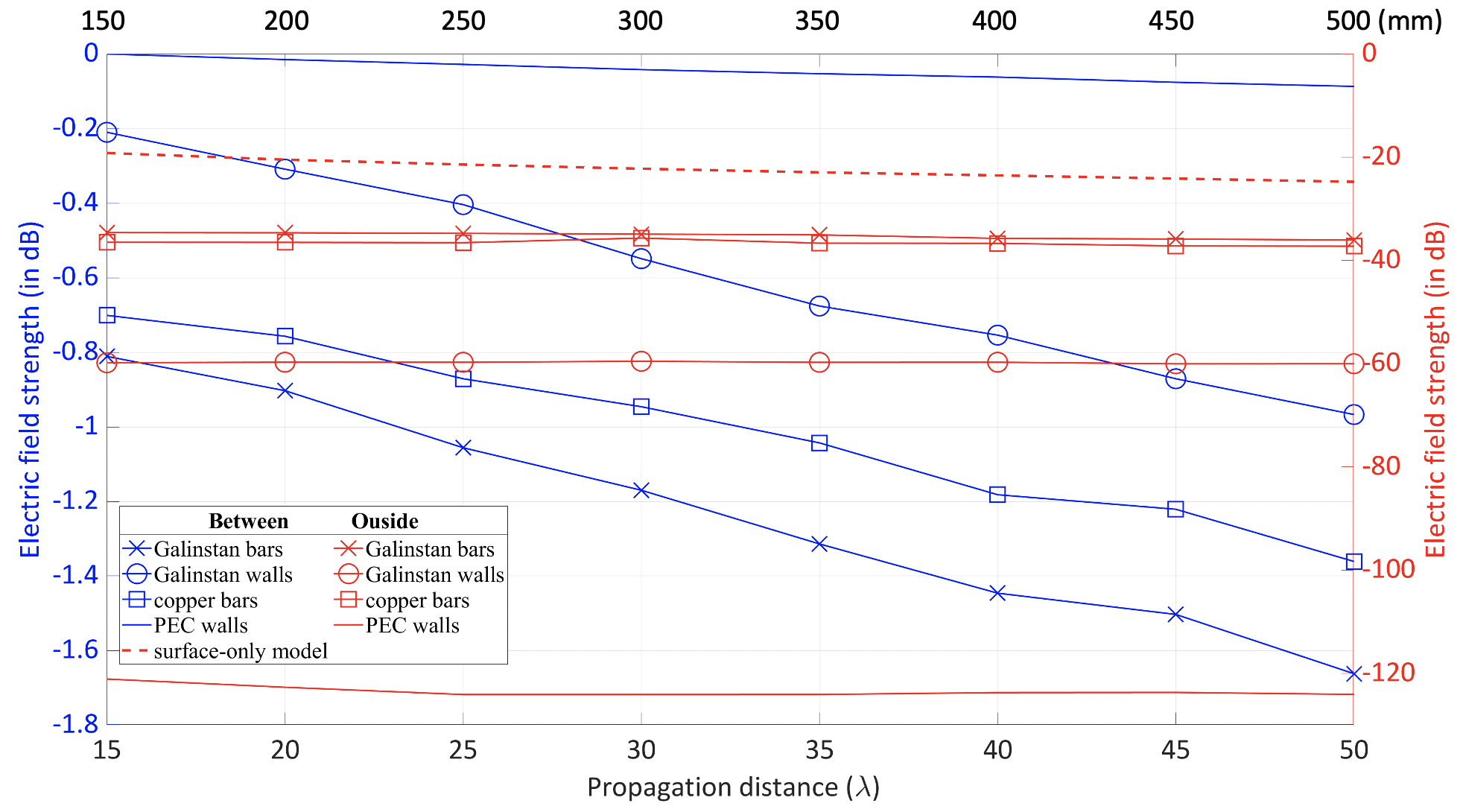}\label{Data_between_and_outside}
}
\caption{Simulation results for (a) the RSWP, and (b) the surface-only model. Also, (c) provides the electric field results at the probe locations in Fig.~\ref{Models} (right) for different models including replacing the bars with PEC walls.}\label{CST_results}
\end{figure}

Results in Fig.~\ref{CST_results}(a) \& (b) are provided for the electric field strengths (in dB) for the RSWP and the surface-only model, respectively. The surface-only model refers to the one without the reconfigurability. As we can see, the electric field for the surface-only model spreads all over the surface and hence has its magnitude dissipated more over distance. By contrast, the RSWP is able to concentrate its propagation along the pathway formed by the fluidic conductive structures. The difference is massive and can be more clearly understood by the results in Fig.~\ref{CST_results}(c) where the results of a couple of other benchmarks are provided, including the ones replacing the Galinstan bars by copper bars or perfect electrical conductor (PEC) walls. In addition, two sets of results are shown in the figure. The blue lines with values according to the left axis correspond to the results at the probes along the pathway supposedly between the bars if present while the red lines accord to the values to the right axis and are the results for the pathway with a tilted angle of $5^\circ$ as depicted in Fig.~\ref{Models}. The results reveal that at $50\lambda$, the RSWP with Galinstan bars has dropped only $0.8{\rm dB}$ while the surface-only model sees a $25{\rm dB}$ loss. We also notice that the difference between using PEC walls and the RSWP with Galinstan is minor, less than $1.5{\rm dB}$ at $50\lambda$. Furthermore, it is observed that at the probe locations outside the pathway, the RSWP records about $40{\rm dB}$ less electric field, meaning that it can block surface waves to unwanted directions although the PEC walls achieve a much less leakage, at $-120{\rm dB}$.

\begin{figure}[]
\centering
\includegraphics[width=9cm]{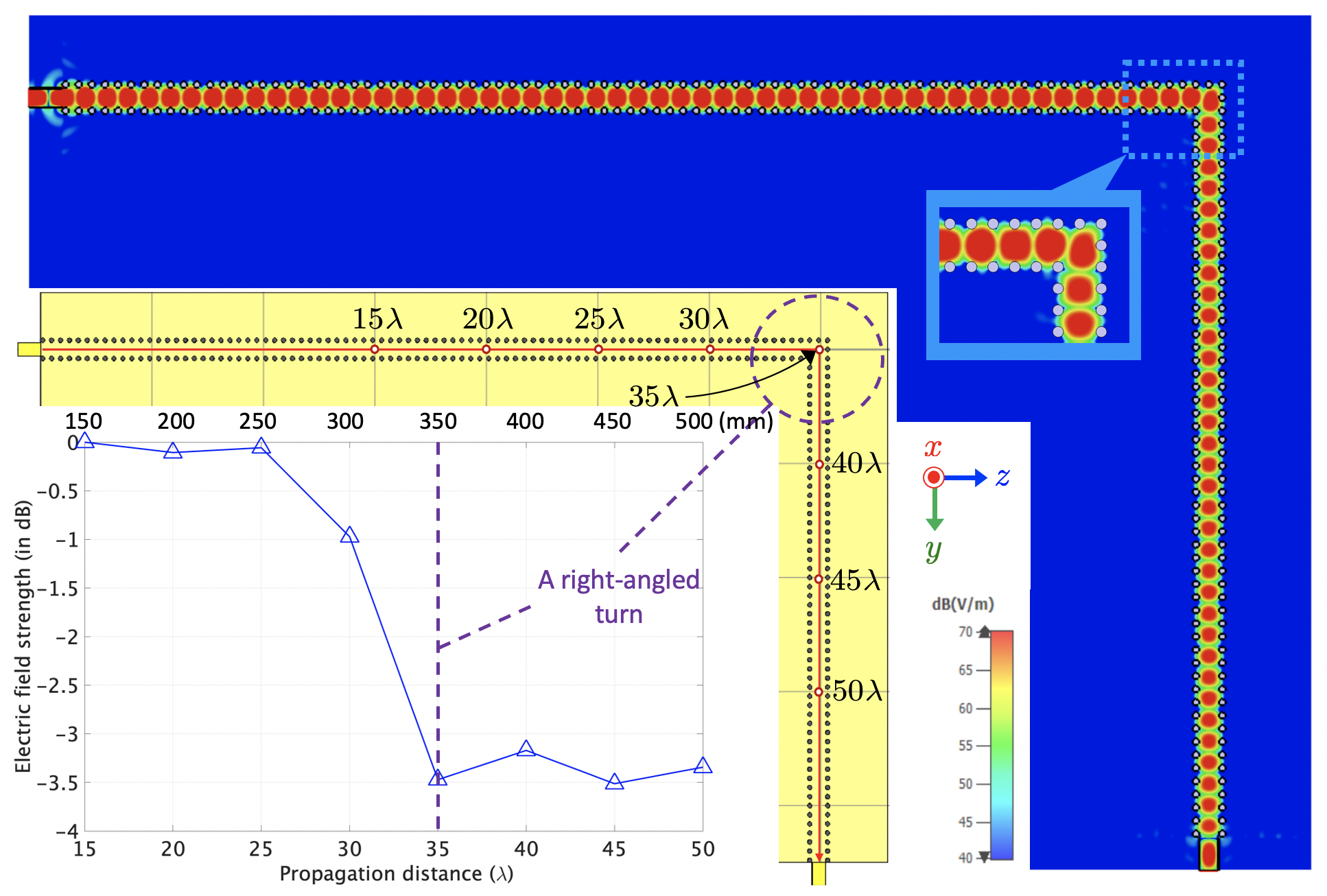}
\caption{Simulation results for an $L$-shaped path created by the RSWP.}\label{Turn_model}
\end{figure}

In Fig.~\ref{Turn_model}, we investigate the feasibility of using the RSWP to form a pathway with sharp turns. An $L$-shaped propagation pathway with a right-angled turn at $35\lambda$ as shown in the figure was considered. The results are consistent with those discussed earlier, able to concentrate the electric field inside the pathway with little leakage outside. The path-loss over distance remains to be negligible ($\approx 0{\rm dB}$ over a distance of $50\lambda$) but the right-angled turn does cause a loss of about $3.5{\rm dB}$.

\section{Conclusion}
This paper presented a new RSWP that can form reconfigurable pathways using an array of cavity bars filled with or without liquid metal on-demand. The proposed platform was evaluated using CST simulations and the results showed great promises of controllable surface waves for energy-efficient and agile communications. With great reconfigurability, the RSWP has the potential to innovate NoC and 6G communications.



\bibliographystyle{IEEEtran}

\begin{thebibliography}{1}
\providecommand{\url}[1]{#1}
\csname url@samestyle\endcsname
\providecommand{\newblock}{\relax}
\providecommand{\bibinfo}[2]{#2}
\providecommand{\BIBentrySTDinterwordspacing}{\spaceskip=0pt\relax}
\providecommand{\BIBentryALTinterwordstretchfactor}{4}
\providecommand{\BIBentryALTinterwordspacing}{\spaceskip=\fontdimen2\font plus
\BIBentryALTinterwordstretchfactor\fontdimen3\font minus
  \fontdimen4\font\relax}
\providecommand{\BIBforeignlanguage}[2]{{%
\expandafter\ifx\csname l@#1\endcsname\relax
\typeout{** WARNING: IEEEtran.bst: No hyphenation pattern has been}%
\typeout{** loaded for the language `#1'. Using the pattern for}%
\typeout{** the default language instead.}%
\else
\language=\csname l@#1\endcsname
\fi
#2}}
\providecommand{\BIBdecl}{\relax}
\BIBdecl

\bibitem{sarkar2017surface}
T.~K. Sarkar, M.~N. Abdallah, M.~Salazar-Palma, and W.~M. Dyab, ``Surface plasmons-polaritons, surface waves, and Zenneck waves: Clarification of the terms and a description of the concepts and their evolution,'' \emph{IEEE Antennas and Propag. Mag.}, vol.~59, no.~3, pp. 77--93, 2017.

\bibitem{agyeman2016resilient}
M.~O. Agyeman, Q.-T. Vien, A.~Ahmadinia, A.~Yakovlev, K.-F. Tong, and T.~Mak, ``A resilient 2-d waveguide communication fabric for hybrid wired-wireless NoC design,'' \emph{IEEE Trans. Parallel and Distributed Sys.}, vol.~28, no.~2, pp. 359--373, 2016.

\bibitem{dai2020reconfigurable}
L. Dai {\em et al.}, ``Reconfigurable intelligent surface-based wireless communications: Antenna design, prototyping, and experimental results,'' {\em IEEE Access}, vol. 8, pp. 45913--45923, 2020.

\bibitem{Wong-2021}
K. K. Wong, K.-F. Tong, Z. Chu and Y. Zhang, ``A vision to smart radio environment: Surface wave communication superhighways,'' {\em IEEE Wireless Commun.}, vol. 28, no. 1, pp. 112--119, Feb. 2021.

\bibitem{gao2018surface}
Z.~Gao {\em et al.}, ``Surface-wave pulse routing around sharp right angles,'' \emph{Physical Review Applied}, vol.~9, no.~4, p. 044019, 2018.

\end{thebibliography}

\end{document}